\documentclass[a4paper, amsfonts, amssymb, amsmath, reprint, showkeys, nofootinbib, twoside,aps,prl]{revtex4-1}
\usepackage[english]{babel}
\usepackage[utf8]{inputenc}
\usepackage{siunitx}  
\usepackage{hhline}
\usepackage{tabularx}
\usepackage{graphicx}
\usepackage[colorinlistoftodos, color=green!40, prependcaption]{todonotes}
\usepackage[pdftex, pdftitle={Article}, pdfauthor={Author}]{hyperref} % For hyperlinks in the PDF
\usepackage{amsthm}
\usepackage{mathtools}
\usepackage{physics}
\usepackage{xcolor}
\usepackage{graphicx}
\usepackage[left=23mm,right=13mm,top=35mm,columnsep=15pt]{geometry} 
\usepackage{adjustbox}
\usepackage{placeins}
\usepackage[T1]{fontenc}
\usepackage{lipsum}
\usepackage{csquotes}

\AtBeginDocument{\RenewCommandCopy\qty\SI} 
\usepackage{nameref}  
\usepackage{hyperref} 

\begin{document}
\title{Zeptosecond \texorpdfstring{$\gamma$}{gamma}-Ray Pulses Generation via FEL-Driven Microbunching and Laser-Compton Scattering
 }
\author{Jinke Xiong\textsuperscript{1,2}}
\author{Hanghua Xu\textsuperscript{3}}
\email[]{xuhh@sari.ac.cn}
\author{Liangliang Ji\textsuperscript{4}}
\author{Chao Feng\textsuperscript{3}}
\email[]{fengc@sari.ac.cn}
\author{Zhentang Zhao\textsuperscript{3}}
\affiliation{\textsuperscript{1} Shanghai Institute of Applied Physics, Chinese Academy of Sciences, Shanghai 201800, China}
\affiliation{\textsuperscript{2} University of Chinese Academy of Sciences, Beijing 100049, China}
\affiliation{\textsuperscript{3} Shanghai Advanced Research Institute, Chinese Academy of Sciences, Shanghai 201210, China}
\affiliation{\textsuperscript{4} State Key Laboratory of High Field Laser Physics, Shanghai Institute of Optics and Fine Mechanics, Chinese Academy of Sciences, Shanghai 201800, China}

\begin{abstract}
We introduce a novel and reliable approach to generate attosecond\textbackslash zeptosecond ($10^{-21} s,zs$), high-energy photon pulse bursts by synergistically exploiting the inherent characteristics of Free-Electron Lasers (FELs) and laser-Compton scattering. Comprehensive numerical simulations confirm the feasibility of the approach. In our study, a 4 $GeV$ electron beam, undergoing the FEL process, emits radiation at 0.7 $nm$ and develops a microbunched structure. These microbunches subsequently interact with a 2 $ps$ laser pulse in a head-on Compton scattering configuration. These results demonstrate that under conditions of well-established electron beam microbunching, the proposed method successfully yields high-brightness $\gamma$-photon pulse bursts with durations shorter than 800 zeptoseconds and exhibiting exceptional signal-to-noise ratios, consistent with theoretical expectations. This technique not only paves the way for probing ultrafast nuclear dynamics but also provides unprecedented tools for quantum science research, enabling explorations of phenomena such as the Quantum Zeno and Anti-Zeno effects.
\end{abstract}

\maketitle

The ultrashort pulse enables signal capture from samples before significant laser-induced changes occur in the target. This concept, known as 'measurement-before-destruction', pioneers new pathways for ultrafast probing of sensitive states and transient processes, allowing us to observe dynamics that would otherwise be obscured by rapid system modifications \cite{RN280}. The required pulse frequency is determined by the characteristic energy of the process under investigation. For instance, molecular dynamics and chemical reactions can be controlled through infrared laser field-driven excitation of several electron volts, while inner-shell electron dynamics can be manipulated using photons with energies ranging from hundreds of eV to several keV \cite{RN155, RN268}. The next frontier lies in time-resolved nuclear dynamics \cite{RN356}, which imposes stringent requirements on both photon energy and pulse duration of ultrashort pulses. Collective nuclear excitation energies span from tens of keV to MeV, while certain photonuclear reactions demand photon energies on the order of tens of MeV or higher \cite{RN267}. The disintegration time of compound nuclei spans from $10^{-19}$ to $10^{-16}$ seconds, with resonance internal conversion occurring on the attosecond time scale \cite{RN272}. Photodisintegration of nuclei \cite{RN274} unfold on the zeptosecond time scale \cite{RN273}, while fission processes and associated phenomena manifest within tens of zeptoseconds \cite{RN263}. Such photon-nucleus interactions in the sub-attosecond and zeptosecond regime offer the unique chance to explore the level density of the compound nucleus far above the yrast line, hitherto an unknown territory, and provide access to nuclei far from the valley of stability \cite{RN263}. 
 \\ \hspace*{1em}Moreover, the ability to manipulate and observe systems on such ultrashort timescales is not limited to nuclear physics research, but also provides new possibilities for exploring the fundamental phenomena of quantum measurement and quantum evolution interaction. Frequent measurements of an unstable quantum state can either inhibit or accelerate its evolution, depending on the measurement frequency relative to the system's intrinsic properties. These phenomena are respectively known as the Quantum Zeno Effect (QZE) \cite{RN351} and the Anti-Zeno Effect (AZE) \cite{RN362}. Both QZE and AZE have been experimentally observed across various platforms \cite{RN375, RN374, RN371, RN372, RN373}, and typical experimental implementations often involve observation time intervals in the microsecond regime. These intriguing quantum phenomena have found applications in diverse fields, including the cooling and purification of quantum systems 
 \cite{RN368}, and the protection of quantum information \cite{RN369},  to preserve quantum coherence and entanglement \cite{RN370}, etc. Exploring even more extreme conditions \cite{RN354, RN376}—utilizing ultra-broadband, ultrashort pulses(attosecond or sub-attosecond duration) at ultrashort time intervals (attosecond timescale)—could provide new insights into QZE and AZE, potentially fundamentally enriching our understanding and enabling novel applications. However, the generation of pulses with such extreme characteristics remains a significant challenge. 
 \\ \hspace*{1em}Current advances in X-ray free-electron lasers (XFELs) are unlocking the potential for intense attosecond X-ray pulse generation and applications. Multiple impressive techniques have successfully demonstrated the generation of X-ray pulses with durations of several hundred attoseconds \cite{RN130, RN136, RN248}. However, from the perspective of FEL physics, achieving significantly higher photon energies (in the MeV regime) and ultrashort pulse durations on the order of tens of attoseconds or even zeptoseconds solely through FEL methodologies faces substantial fundamental limitations and would require prohibitive resources. Exploring novel methodologies beyond established FEL principles is essential for accessing the zeptosecond timescale and generating $\gamma$-ray pulses. 
\\ \hspace*{1em}Here, we propose a novel and reliable method that combines FEL technology with laser Compton scattering to generate ultrashort (hundreds of zeptosecond) pulses with extremely high photon energies (ranging from several MeV to hundreds of MeV). This approach exploits the established efficacy of Compton scattering as a premier source of high-quality gamma rays while circumventing the pulse broadening typically associated with slippage effects in FEL pulse generation.

  %I believe leaving the sections in separate files is more organized, change it if you desire 

The scheme of generating high-brightness, high-energy, and ultrashort photon pulses through laser-relativistic electron beam interaction was experimentally validated in 1995 \cite{RN275}. In this pioneering work, laser-electron scattering was implemented at 90° geometry to constrain the interaction duration, successfully demonstrating the generation of femtosecond-scale hard X-ray(0.4 angstrom) pulses. For head-on collision configuration, the pulse duration of $\gamma$-rays generated through laser-relativistic electron Compton scattering is characterized by $\sigma_{\tau}=\sqrt{\sigma_{e}^2+\sigma_{L}^2/16\gamma^4}\sim\sigma_e+\sigma_L/4\gamma^2$, where $\sigma_e$ denotes the electron bunch length, $\sigma_L$ represents the laser pulse duration, and $\gamma$ corresponds to the relativistic factor of the electron beam. Following this fundamental principle, when $\gamma$ reaches a sufficiently large value, the scattered photon pulse duration asymptotically approaches the electron bunch length. Although generating such an individual ultra-narrow, high-density electron slice encounters substantial technical challenges, the FEL gain process inherently facilitates the formation of a train of ultra-narrow, high-density electron microbunches. 

A defining characteristic of FELs is the reciprocal interaction between the emitted radiation and the electron beam, leading to a substantial enhancement in both radiation intensity and coherence compared to synchrotron radiation sources based on storage rings. This enhancement is driven, largely, by the formation of periodic density modulations within the electron beam, a phenomenon known as microbunching. These modulations occur at the resonant wavelength, fostering constructive interference and a corresponding increase in longitudinal coherence. Microbunching is a fundamental consequence of the interaction between the relativistic electron beam and the co-propagating radiation field within the periodic magnetic field of the undulator. This interaction imprints a fine structure onto the electron beam’s longitudinal phase space, characterized by a periodicity matching the fundamental wavelength of the undulator radiation. Microscopically, this self-organization arises from the ponderomotive potential generated by the superposition of the undulator and radiation fields. Electrons residing at negative ponderomotive phases experience a net energy gain from the radiation field, while those at positive phases transfer energy to the field. The dispersive properties of the electron motion within the undulator translate this energy modulation into a spatial density modulation, manifesting as the periodic microbunch structure. 

Theoretically, we can generate an ultrashort $\gamma$-ray pulse train through Compton scattering between a laser beam and the microbunched electron beam produced in FEL processes e.g. self-amplified spontaneous emission (SASE) or seeded FEL \cite{RN278, RN277, RN103, RN279}, as shown in Fig.1. This pulse train inherits the characteristics of the microbunch structure of electron beam, featuring a periodicity equal to the FEL radiation wavelength and an even shorter pulse duration. The duration of individual pulse is determined by the FEL radiation wavelength, while the pulse signal-to-noise ratio is governed by the degree of electron microbunching. This methodology yields attosecond $\gamma$-ray pulse trains using tens-of-$nm$ FEL radiation, scaling to the zeptosecond regime when driven by \AA-wavelength FELs.

\begin{figure}[htbp]
    \centering
    \includegraphics[width=1\linewidth]{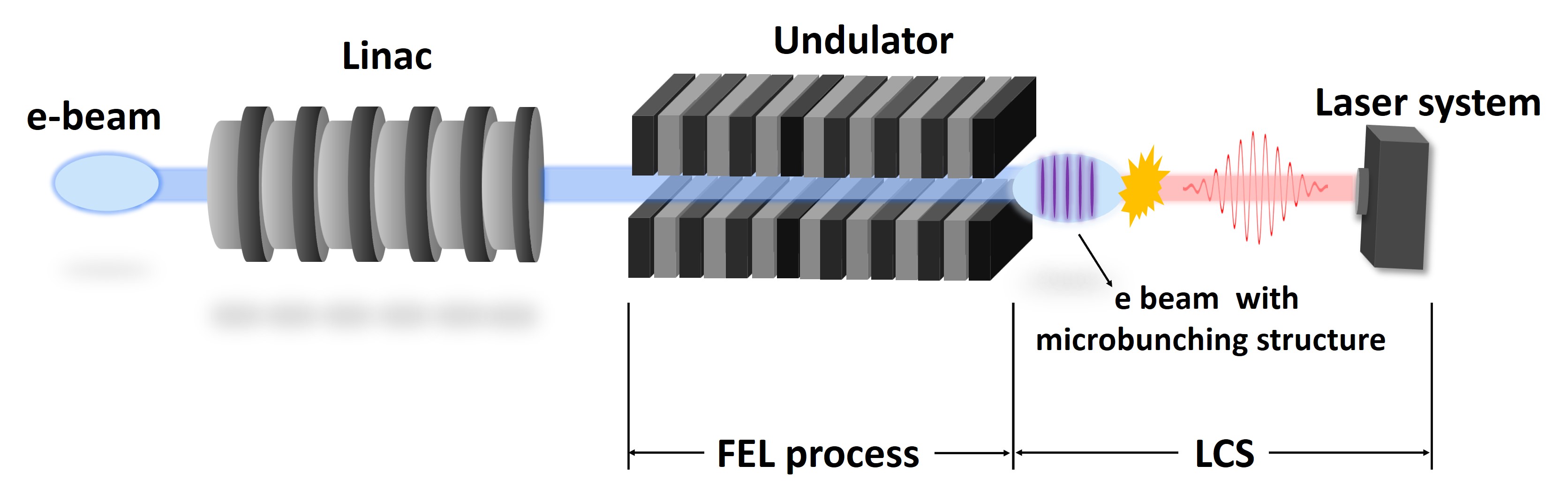}
    \caption{Schematic illustration of the method for generating ultrashort pulse trains}
    \label{fig:fig1}
\end{figure}

The electron density distribution within each microbunch is quantitatively characterized by the bunching factor at fundamental wavelength $b_1=\langle e^{-i\theta_j}\rangle_{\Delta}$, where $\theta_j$ represents the ponderomotive phase of an electron within the FEL slice \cite{RN245, RN365}. The bunching factor $b_1$ is the ensemble average of the complex exponential function, evaluated over the phases of all electrons within a temporal slice of the FEL interaction region. Specifically, a larger bunching factor magnitude (ranging from 0 to 1) corresponds to higher charge density within the microbunch. When considering the microbunch as a pulse signal, the bunching factor magnitude substantially reflects the signal-to-noise ratio, which is subsequently projected onto the scattered photon distribution during the Compton scattering process. The laser Compton scattering process can be analogously viewed as electrons traversing a laser-undulator \cite{RN365, RN275}, where electron slices with higher charge density generate correspondingly higher radiation power. 

Currently operational and under-construction hard XFEL facilities predominantly operate in the SASE mode, capable of generating femtosecond X-ray pulses with gigawatt peak power. During the SASE process, electron beams traverse through an extended undulator section, where the spontaneous undulator radiation noise undergoes exponential amplification until reaching saturation. Theoretically, the SASE operational mode exhibits wavelength-independent characteristics, enabling its functionality across arbitrary spectral regions. This wavelength flexibility inherently suggests the potential for achieving finer microbunching structures.

Subsequently, we investigate the feasibility of generating ultrashort, high-energy $\gamma$-ray pulse trains through the collision between laser pulses and electron beams exhibiting microbunch structures produced by the SASE FEL 
process. The formation of the microbunching is numerically simulated using Genesis 1.3 
 \cite{RN282, RN283}with the typical parameters of an X-ray FEL, as shown in Table.1. 

\begin{table}[htbp] % [htbp] 是可选的, 用于控制表格的浮动位置 
  \noindent
  \centering
  
  \caption{Main parameters used in FEL simulations} % 表格标题
  \label{tab:parameters} % 表格标签 
   \begin{tabularx}{\linewidth}{X >{\centering\arraybackslash}X}  %改X 为 l c
    \hhline{==}% 顶部横线
    Parameters & {Value} \\ % 表头 (注意 Value 使用花括号, 因为 siunitx 需要)
    \hline % 中间横线
    Energy            & 4 \si{\giga\electronvolt}  \\ % 使用 siunitx 排版单位
    Average current   & 3.5 \si{\kilo\ampere}     \\
    Emittance         & 0.4 \si{\milli\meter\milli\radian} \\
    Energy spread     & 0.01 \si{\percent}        \\
    Radiation wavelength & 0.7 \si{\nano\meter}    \\
    Undulator period  & 3.5 \si{\centi\meter}   \\
     \hhline{==} % 中间横线
  \end{tabularx}
\end{table}

In this example, we set the radiation wavelength of the FEL process to 0.7 $nm$. Figure 2 shows the longitudinal spatial distribution of microbunches generated by the FEL numerical simulation, along with the corresponding current profile.

\begin{figure}[htbp]
    \centering
    \includegraphics[width=1\linewidth]{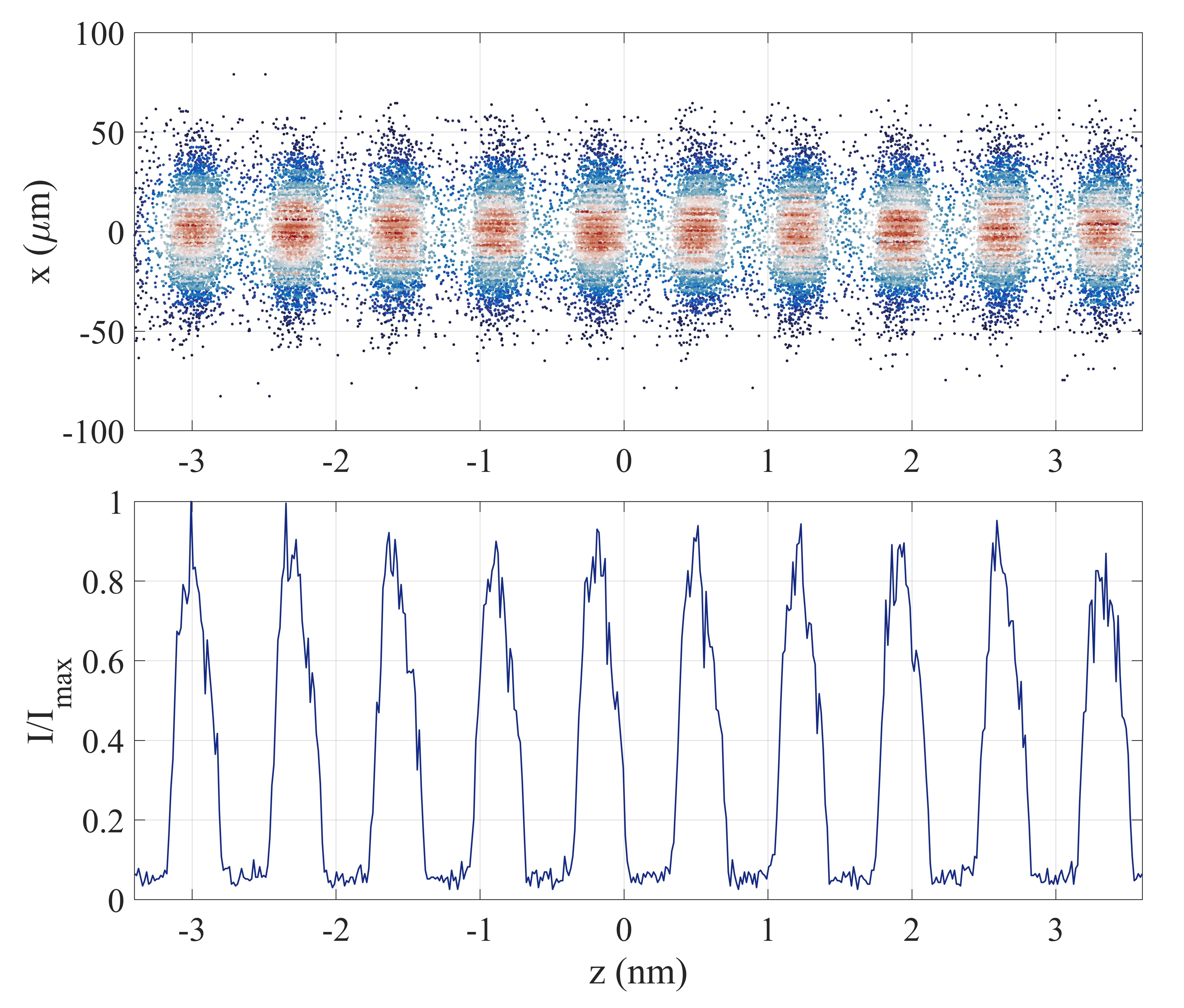}
    \caption{Spatial distribution and current profile of the electron microbunches at a bunching factor of $\sim$ 0.6, where $I_{max}$ represents the peak current. }
    \label{fig:fig2}
\end{figure}

The scattered  sub-attosecond $\gamma$-ray pulses produced via Compton scattering are emitted as a highly collimated beam with a divergence angle of approximately l/$\gamma$ (0.1277 mrad in this case), where $\gamma$ denotes the relativistic factor of the electrons. In the linear Compton scattering regime, the energy of scattered photons can be expressed as \cite{RN160, RN275}:
\begin{equation}
   E=E_{laser}2\gamma\frac{1-cos\phi}{1+\gamma^2\theta^2}
\end{equation}
where $E_{laser}=hc/\lambda_{laser}$ represents the laser photon energy ($h$ denotes Planck's constant, $c$ is the speed of light in vacuum, and $\lambda_{laser}$ is the wavelength of the incident laser), $\phi$ represents the interaction angle between electrons and photons ($\phi$=$\pi$ in our case), and $\theta$ denotes the emission angle of $\gamma$-rays relative to the electron trajectory. From this expression, it is evident that the scattered photon energy can be tuned by adjusting either the incident laser wavelength or the interaction angle between the laser and electron beams, while keeping the electron energy constant, potentially reaching a maximum value of $4\gamma^2$ times the laser photon energy. 
The total yield of scattered photons can be estimated, while neglecting quantum recoil effects. For the head-on collision configuration, the yield can be approximated as: 
\begin{equation}
    N_{\gamma}= \frac{\sigma N_e N_{laser}}
    {
    2\pi \sqrt{\sigma_{e\ x}^2+\sigma_{laser\ x}^2} 
    \sqrt{\sigma_{e\ y}^2+\sigma_{laser\ y}^2}
    }
\end{equation}
where $N_e, N_{laser}$ represent the number of electrons and photons in the beams, respectively, $\sigma$=0.665 $barn$ denotes the total scattering cross-section, and $\sqrt{\sigma_{e\ x}^2+\sigma_{laser\ x}^2} \sqrt{\sigma_{e\ y}^2+\sigma_{laser\ y}^2}$ is the effective interaction area, given by the transverse size convolution of the electron and laser beams. This equation indicates that the photon yield scales linearly with the number of interacting electrons and photons and inversely with the effective interaction area. In this example, with a microbunch charge of approximately 8 $fC$ and a laser pulse energy of 0.2946 $J$, the estimated photon yield within a solid angle of 1/$\gamma$ is approximately 3000 photons. 

We employed CAIN, a well-established Monte Carlo code \cite{RN276, RN252}, for simulating beam-beam interactions, to numerically model the laser Compton scattering process. CAIN has been extensively used in design studies of inverse Compton scattering sources. 

The parameters of an intense and ultra-short infrared laser \cite{RN358} have been adopted in these simulations. The pulse width can be as short as 3 $ps$ and the peak power reaches 15 $TW$(45$J$). The parameters of the laser are listed in Table. 2 in this example. 

\begin{table}[htbp] % [htbp] 是可选的, 用于控制表格的浮动位置 
  \noindent
  \centering
  \caption{Laser parameters} % 表格标题
  \label{tab:2} % 
   \begin{tabularx}{\linewidth}{X >{\centering\arraybackslash}X}  %改X 为 l c
   \hhline{==} 
    Parameters & {Value} \\ 
    \hline 
    Power intensity $P$           & $1.5 \times 10^{19} \si{\watt\per\meter\squared}$  \\ % 使用 siunitx 排版单位
    RMS beam size at focus $\omega_{0}$   & 
    25 \si{\micro\meter}    \\
    wavelength $\lambda_{laser}$        & 
    9 \si{\micro\meter} \\
    RMS duration $\sigma_{laser}$     & 
    2 \si{\pico\second}        \\
    Pulse energy & 0.2946 \si{\joule}    \\
    \hhline{==} 
    \end{tabularx}
\end{table}

Numerical simulations of the laser-Compton scattering process are grounded in the Quantum Electrodynamics (QED) framework. In the presented scenario, utilizing circularly polarized laser light with a laser intensity parameter $ \xi=\frac{e \sqrt{-A^{\mu}A_{\mu}}}{m_e} \sim \frac{\lambda_{laser}}{2\pi m_e}\sqrt{\mu_{0}cP} $ of 0.21, where nonlinear effects remain negligible. The temporal broadening of the radiation pulse, caused by the incident laser pulse duration, is orders of magnitude shorter than the characteristic microbunch length of 778 $zs$ (FWHM). We use a heuristic approximation here, estimating the microbunch’s FWHM width to be one-third of the radiated wavelength.

Figure.3. illustrates the numerical simulation results, depicting the temporal distribution of the scattered gamma photons and the FWHM for each pulse. The results indicate that with a bunching factor of approximately 0.6, the pulse train exhibits a favorable signal-to-noise ratio, and each individual pulse has a duration of 780 ± 20 $zs$, consistent with the theoretical estimate of 780 $zs$. The inter pulse separation is approximately 2.33 attoseconds (0.7 $nm$). From fundamental FEL principles, the spacing between consecutive microbunches remains fixed at the fundamental wavelength of the undulator radiation, ensuring minimal jitter in the peak-to-peak separation within the pulse train. The numerical simulation results presented in Fig.4.(right) also confirm this. Fig.3.(right) illustrates the spectral distribution of the pulse train, representing the superposition of the energy spectra from each individual pulse. The near-independent nature of the interaction between each micro-bunch and the laser pulse allows for the reasonable assumption that the energy spectrum of gamma photons originating from each interaction is consistent. Consequently, the spectral bandwidth of the Compton scattering radiation pulses can be controlled by employing a collimation aperture \cite{RN160}. This tunability stems from the one-to-one mapping between the emission angle $\phi$ and energy of the scattered photons. However, this approach to spectral bandwidth control inevitably sacrifices gamma-ray photon yield. 

\begin{figure*}[t]
    \centering
    \includegraphics[width=0.97\textwidth]{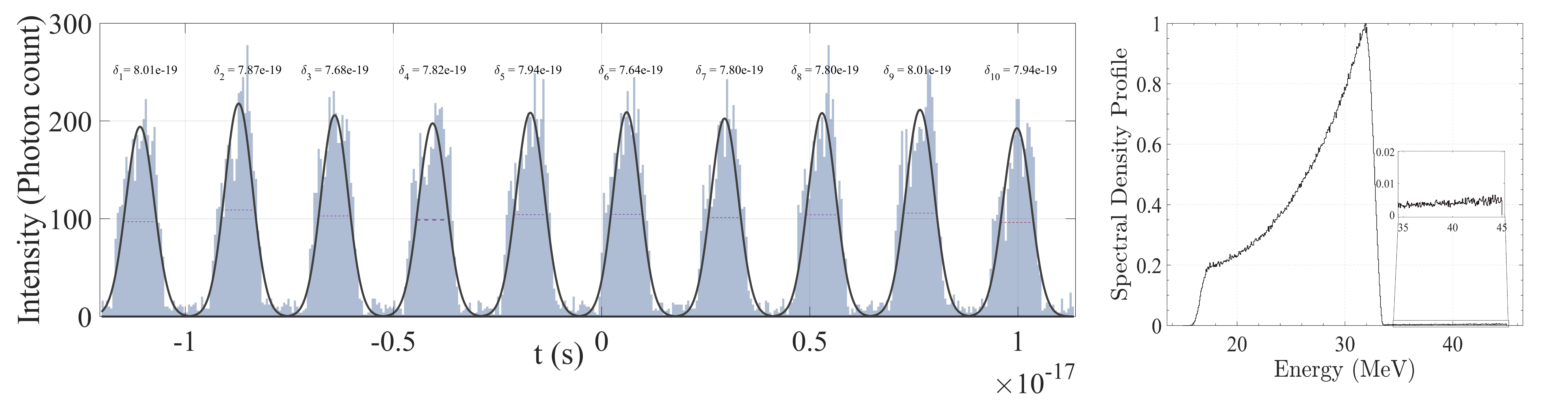}
    \caption{(left)The temporal profile of the scattered photons within a scattering angle $\theta < 1/\gamma$ correlates with the microbunching structure depicted in Fig.2. The $\delta$ ($FWHM$) of each individual pulse is also shown. The $FWHM$ calculated using the formula $\sigma_{\tau}=\sqrt{\sigma_e^2+\sigma_{laser}^2/16\gamma^4}$ is approximately 778.6$zs$ ($\sim2.355\sigma_{\tau}$ ). The vertical axis represents the photon count. (right)Pulse train energy spectral density distribution}
    \label{fig:fig3}
    
\end{figure*}

 \begin{figure}[htbp]
     \centering
      \includegraphics[width=0.49\linewidth]{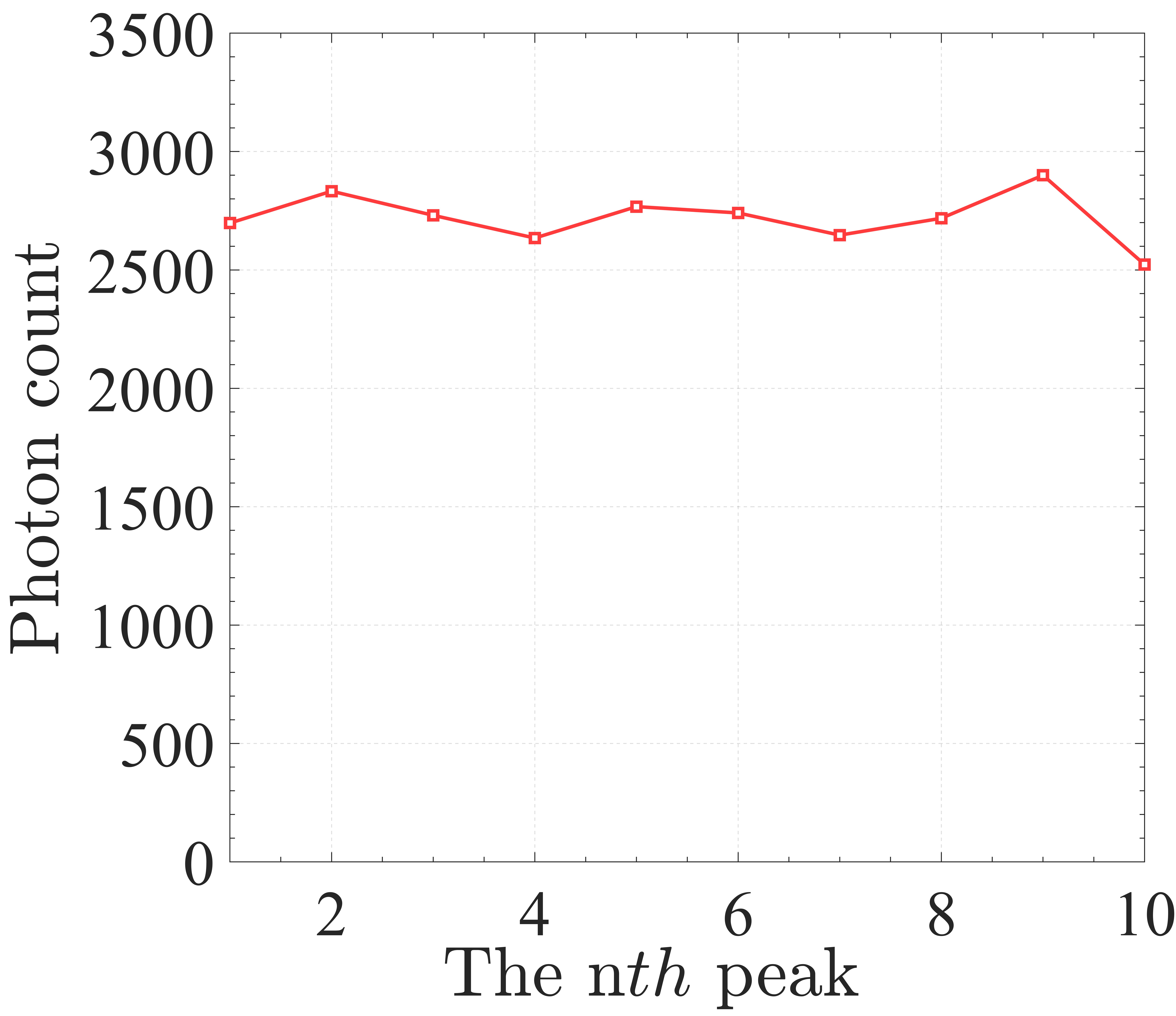}
     \hfill 
      \includegraphics[width=0.49\linewidth]{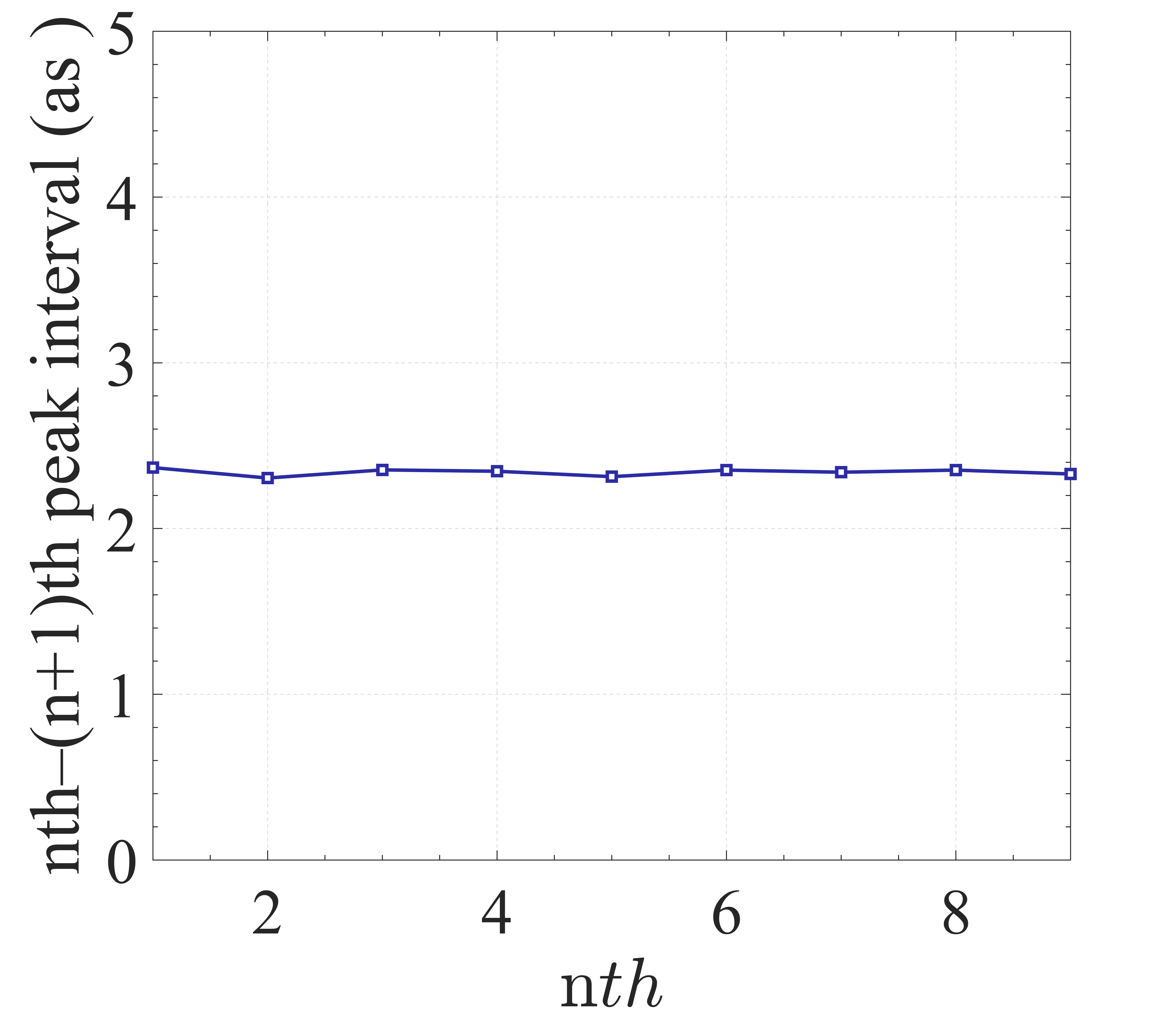}
     \caption{(left) Number of photons within the FWHM of the nth peak.  (right) Interval between n$th$ and (n+1)$th$ peaks. }
       \label{fig5a5b}
\end{figure}

A substantial enhancement of the scattered gamma-ray photon yield can be achieved, while maintaining a constant $\xi$, by increasing the incident laser pulse duration. Crucially, this increase in laser pulse duration negligibly affects the duration of individual pulses within the generated pulse train.

In electron beams with non-uniform current distributions, variations in the charge within individual microbunches contribute to fluctuations in the gamma-ray photon yield from pulse to pulse. Crucially, the signal-to-noise ratio of the gamma-ray pulse train is directly correlated with the distribution of the bunching factor within the electron bunch. An additional obstacle arises at shorter FEL wavelengths, where the charge captured within each microbunch decreases (strongly correlated with the electron beam current), thus reducing the photon yield of individual pulses. The propagation of such ultrashort pulses also poses a challenge. As the propagation distance $L$ increases, pulse duration broadening, $\Delta\tau$, becomes increasingly pronounced, following the relationship $\Delta\tau\propto L\theta_{RMS}^2/2c$, reaching the attosecond scale. While typically negligible when the initial pulse duration is larger, this broadening effect becomes prominent in our scenario(sub-attosecond regime), surpassing the initial pulse duration and subsequently compromising the pulse train's signal-to-noise ratio. Currently, mitigation strategies, apart from increasing the electron beam energy, are limited to restricting the gamma-ray beam propagation distance or employing collimation apertures to control divergence. As previously discussed, this measures introduce a trade-off: although beam collimation can narrow the radiation pulse bandwidth, it is accompanied by a reduction in the gamma photon yield. Regrettably, in the example presented, even with the constraint imposed by a collimation aperture, the sample target position remains restricted to a mere 1-meter range from the source. 

The presented example serves as a proof-of-principle demonstration of the feasibility of generating zeptosecond-scale pulses by combining FEL technology with laser Compton scattering. 
Our results demonstrate that with a 4 $GeV$ electron beam and an FEL radiating at 0.7 $nm$, a relatively stable pulse train with a favorable signal-to-noise ratio, tunable control of scattered photon energy and a pulse duration of approximately 800 $zs$ can be generated. 
Remarkably, even within the linear Compton scattering regime, individual pulses contain approximately 3000 photons within a solid angle of $\theta<1/\gamma$. Furthermore, operating at longer radiation wavelengths, which enables lower electron beam energy, would increase the trapped charge within each microbunch and enhance the bunching factor, consequently leading to a significant improvement in the per-pulse photon yield, the signal-to-noise ratio and significantly curtailing the negative impact of propagation. For example, with this scheme operating at an electron beam energy of 1.5 $GeV$ and an FEL radiation wavelength of 10 $nm$, the generated approximately 10 attoseconds pulse train exhibits slight sensitivity to propagation effects concerning pulse duration and signal-to-noise ratio. 

Additionally, the $\gamma$-ray pulses maintain precise synchronization with the optical laser pulses, which is essential for pump-probe applications. The pulse train produced via this method possesses a significantly high frequency (pulses per second). This high frequency is reminiscent of the Quantum Zeno Effect and Anti-Zeno Effect. The proposed scheme can easily provide measurement frequencies on the order of $10^{17}$ $Hz$,  potentially offering unprecedented perspectives for experimental investigations of the QZE and AZE. Specifically, theoretical predictions suggest that the AZE could be observed in nuclear $\beta$-decay by perturbing the decaying nucleus with a high-frequency, broadband $\gamma$-ray source of this nature \cite{RN363, RN354, RN362}, thus opening up possibilities for controlling a range of decay processes.

\section*{Acknowledgements} \label{sec:acknowledgements}
The authors thank Chang Xu, Hongwei Wang, and Gongtao Fan for helpful discussions. This work was supported by the CAS Project for Young Scientists in Basic Research (YSBR-115) and National Natural Science Foundation of China (12435011).

\bibliography{./reference}

\end{document}